\documentclass[manuscript]{acmart}

\AtBeginDocument{%
  }

\usepackage{dirtytalk}    
\usepackage{graphicx} 
\usepackage[
    colorinlistoftodos,
    textsize=footnotesize,
        ]{todonotes}

\definecolor{darkgreen}{rgb}{0.0, 0.5, 0.0}

\definecolor{lightyellow}{HTML}{FFE699}
\definecolor{red_revision}{HTML}{FF0000}

\title{The Effect of Education in Prompt Engineering: Evidence from Journalists}

\author{Amirsiavosh Bashardoust\textsuperscript{*}}
\email{amirsiavosh.bashardoust@unil.ch}
\affiliation{%
  \institution{University of Lausanne}
  \city{Lausanne}
  \country{Switzerland}
}

\author{Yuanjun Feng\textsuperscript{*}}
\affiliation{%
  \institution{University of Lausanne}
  \city{Lausanne}
  \country{Switzerland}
}

\author{Dominique Geissler\textsuperscript{*}}
\email{d.geissler@lmu.de}
\affiliation{%
  \institution{Munich Center for Machine Learning, LMU Munich}
    \city{Munich}
  \country{Germany}
}

\author{Stefan Feuerriegel}
\email{feuerriegel@lmu.de}
\affiliation{%
  \institution{Munich Center for Machine Learning, LMU Munich}
    \city{Munich}
  \country{Germany}
}

\author{Yash Raj Shrestha}
\email{yashraj.shrestha@unil.ch}
\affiliation{%
  \institution{University of Lausanne}
  \city{Lausanne}
  \country{Switzerland}
}

\begin{document}
 \renewcommand{\thefootnote}{\fnsymbol{footnote}}
 \footnotetext[1]{These authors contributed equally to this work.}

\settopmatter{printacmref=false}
\setcopyright{none}
\renewcommand{\shortauthors}{Bashardoust, Feng, Geissler, Feuerriegel and Shrestha}

\begin{abstract}

Large language models (LLMs) are increasingly used in daily work. In this paper, we analyze whether training in prompt engineering can improve the interactions of users with LLMs. For this, we conducted a field experiment where we asked journalists to write short texts before and after training in prompt engineering. We then analyzed the effect of training on three dimensions: (1)~the user experience of journalists when interacting with LLMs, (2)~the accuracy of the texts (assessed by a domain expert), and (3)~the reader perception, such as clarity, engagement, and other text quality dimensions (assessed by non-expert readers). Our results show: (1)~Our training improved the perceived expertise of journalists but also decreased the perceived helpfulness of LLM use. (2)~The effect on accuracy varied by the difficulty of the task. (3)~There is a mixed impact of training on reader perception across different text quality dimensions.

\end{abstract}



\maketitle

\section{Introduction}
\label{sec:introduction}


Generative AI such as large language models (LLMs) are increasingly integrated into daily work \cite{Littman.2021.AIreport, Feuerriegel.2023}. In a recent survey, 26\% of workers reported that they already use LLMs several times each week \cite{Beauchene.2023} and that LLMs have a large impact on their productivity \cite{Noy.2023, Brynjolfsson.2023, Li.2024}. For example, workers who use LLMs are more efficient and experience more joy during work \cite{Noy.2023}. Hence, it is not surprising that 86\% of employees also believe they need training to fully leverage the potential of LLMs in their workplace \cite{Beauchene.2023}. 

The quality of LLM outputs depends -- to a large extent -- on the quality of prompts, meaning the user input that the LLM receives \cite{Atreja.2024}. For example, poorly structured prompts tend to result in vague, incorrect, or irrelevant output \cite{Lin.2024}. Output quality can be assessed through different lenses, i.e., through domain experts and non-expert readers \cite{McGuire.2024}. Moreover, the quality of prompts also shapes the overall {experience of users}. For example, research has found that novice users have difficulties designing effective prompts \cite{ZamfirescuPereira.2023}. This can lead to dissatisfaction among users \cite{Kim.2024} and reduce the perceived helpfulness of LLMs, as, in particular, novice users often experience increased frustration and struggle to adjust prompts effectively to improve their outcomes \cite{Kim.2024}.


Effective prompt writing, commonly referred to as \emph{prompt engineering}, is widely regarded as an essential skill when using LLMs \cite{Yang.2024}. As we hypothesize later, this can eventually improve the {output quality} of LLMs and the {user experience}. Effective prompt engineering relies on understanding the capabilities and limitations of LLMs and writing prompts in such a way that precise, relevant, and accurate information is produced. So far, different strategies for prompt engineering are recommended, for which common examples include asking the LLM to use chain-of-thought reasoning \cite{Wei.2022} or to adopt the role of a specific persona \cite{White.2023}. Yet, prompt engineering is a new skill that needs to be learned to be leveraged at full potential \cite{ZamfirescuPereira.2023}. This raises the question of whether prompt engineering training can help improve both the output quality and the user experience when interacting with LLMs. 

\begin{figure}
    \centering
    \includegraphics[width=\linewidth]{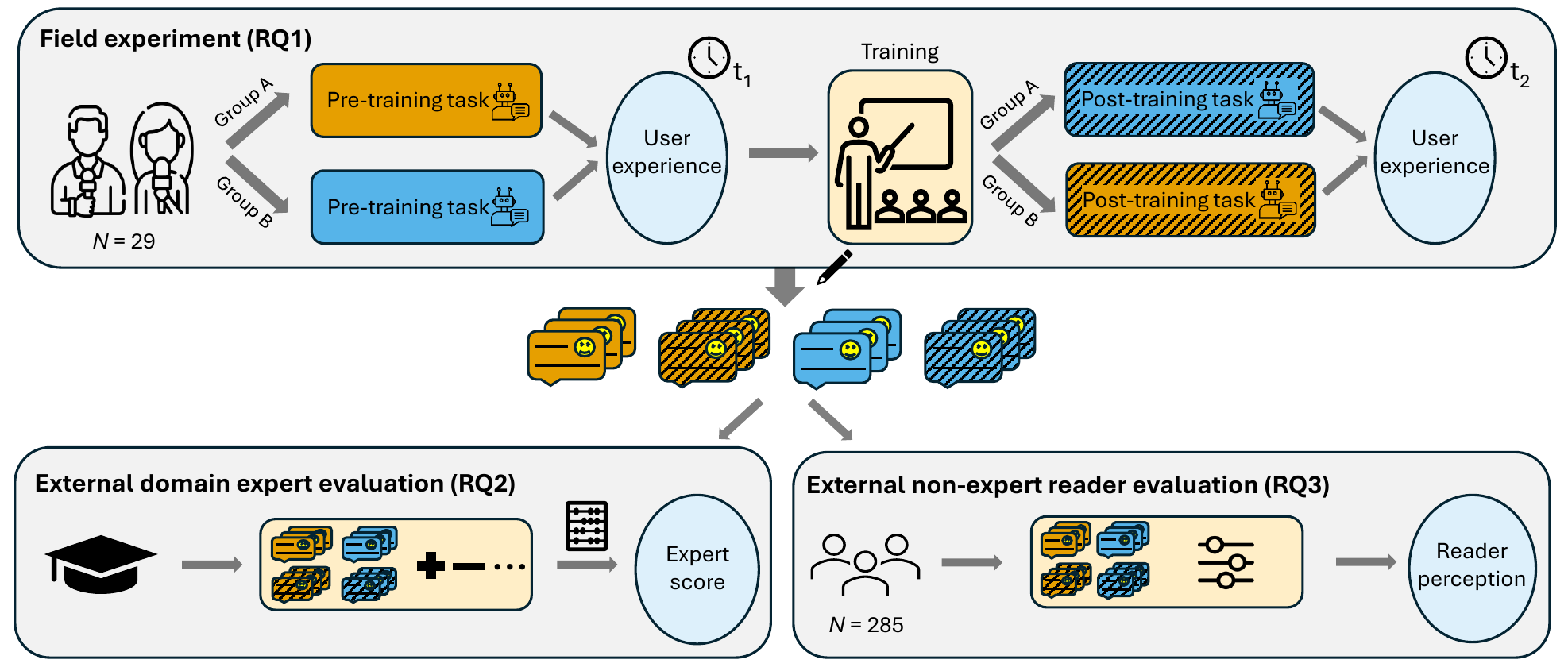}
    \caption{Overview of our study process.}
    \label{fig:overview_graphic}
    \Description{Flowchart illustrating the flow of our study.  First, two groups of participants, labeled Group A and Group B, complete initial writing tasks together with an LLM before undergoing training. After the training, both groups perform similar post-training tasks. The participants' user experiences are measured both before and after training. In the second phase, an external expert evaluates the results of these writing tasks and assigns scores. Finally, in the third phase, external non-expert readers assess the same data to gauge their perception. The study aims to compare the outcomes across these different evaluations to analyze the effect of prompt engineering training.}
\end{figure}

In this paper, we analyze whether prompt engineering training can improve the user experience and output quality of users interacting with LLMs. For this, we conducted a field experiment with a counter-balanced measures design where $N=29$ participants receive in-person training in prompt engineering.\footnote{Code and data for this project are available publicly at: \href{https://github.com/vosh-96/The-Effect-of-Education-in-Prompt-Engineering-Evidence-from-Journalists}{https://github.com/vosh-96/The-Effect-of-Education-in-Prompt-Engineering-Evidence-from-Journalists}.} We performed our experiment with professional journalists whom we asked to write short social media posts about scientific research articles with the support of an LLM both before and after training  (see Figure~\ref{fig:overview_graphic} for an overview of our study flow). We analyze the effect of training on three dimensions: (1)~We asked the participants of the training about their perceived expertise and perceived helpfulness of the LLM during the field experiment. This way, we can analyze the effect of training on user experience. (2)~We asked a domain expert to rate the accuracy of the texts written during training, with which we measured output quality through an expert's lens. (3)~We also measure output quality through the lens of non-expert readers by asking about their perception of the texts along multiple dimensions, such as clarity and engagement.  

For our study, we chose journalists to perform the prompt engineering training with. This is because the daily tasks of journalists are representative of knowledge work, where workers have to demonstrate complex problem-solving abilities. In addition, journalism is a field that is at high risk of exposure to Generative AI \cite{Felten.2023}. The use of Generative AI in journalism is rising rapidly as Generative AI is increasingly leveraged by journalists in various tasks such as writing texts and multi-media content creation \cite{FAZ.2024}. For example, the \emph{Los Angeles Times} already uses Generative AI to automatically generate reports on seismic activity in real-time \cite{LosAngelesTimes.2019}. It is thus important for journalists to adapt to new AI technologies to stay relevant in a rapidly changing media landscape which raises the need to understand whether journalists can be trained in prompt engineering.

\section{Related work}
\label{sec:related_works}

We review two literature streams that are particularly relevant to our study, namely, (i) literature aimed at improving AI literacy through training and (ii) literature on prompt engineering that provides strategies and behavioral evidence.

\subsection{Training aimed at AI literacy}

AI literacy refers to \say{a set of competencies that enables individuals to critically evaluate AI technologies; communicate and collaborate effectively with AI; and use AI as a tool online, at home, and in the workplace} \cite{Long.2020}. Given the increasing integration of AI into professional environments, AI literacy plays a crucial role in preparing workers to effectively interact with AI technologies \cite{Pinski.2023}. As a result, there is a growing demand for AI literacy training programs that equip individuals with the necessary skills to make effective use of AI technology \cite{Ng.2021}. To achieve this goal, UNESCO developed a curriculum that defines essential learning domains for AI literacy training. These include AI fundamentals, ethical considerations, societal impacts, and AI application and development \cite{UNESCO.2022}. Our work advances this effort by extending the understanding of how to effectively enhance users' practical skills in prompt engineering, resulting in improved output quality when interacting with LLMs.

Previous work has developed various trainings for AI literacy. For example, \citet{Kandlhofer.2016} designed AI literacy trainings for different age groups from kindergarten to university and found that the trainings improved AI literacy. Another work \cite{Markus.2024} developed online trainings, where participants were taught how AI technologies function to improve their understanding of intelligent voice assistants. The work found that the trainings (i)~ led to an improved understanding of the potential and risks of intelligent voice assistants, (ii)~promoted effective interactions with the voice assistants, and (iii)~elicited user interest in exploring the technology. \citet{Moore.2022} designed a 1-hour interactive module where participants were trained in recognizing online misinformation. The authors found that the training improves truth discernment, especially for older adults. \citet{Theophilou.2023} studied how prompt engineering training can help users understand the opportunities and limits of LLMs. The authors found that training reduced negative sentiments toward LLMs and that students had a better understanding of the limitations of LLMs. Yet, unlike our work, this research focuses on understanding AI limitations and does not consider the output quality of LLMs. Still, empirical evidence is missing on how prompt engineering training can affect task-specific output quality, particularly in professional environments. 

\subsection{Prompt engineering: definition, strategies, and behavioral evidence}

The quality of the LLM output highly depends on the quality of the user input (i.e., the prompt) \cite{Atreja.2024}. Some general guidelines on how to effectively prompt LLMs have been developed, which include: (1)~guiding the model to solutions, (2)~adding relevant context, (3)~being explicit in the instructions, (4)~asking for lots of options, (5)~giving examples of good answers, and (6)~breaking complex tasks into subtasks \cite{Lin.2024, OpenAI.2024, Mesko.2023, White.2023}. 

Beyond simple guidelines, several, more advanced strategies for \emph{prompt engineering} have been identified that generally tend to obtain high-quality outputs from the LLM \cite{Liu.2023}. One strategy is to ask the model to use \emph{chain-of-thought reasoning}. Chain-of-thought describes a series of intermediate natural language reasoning steps that lead to the final output \cite{Wei.2022}. This allows LLMs to decompose multi-step problems into solvable, intermediate steps, which, in turn, can positively affect the accuracy and relevance of the output. Furthermore, chain-of-thought prompting enables users to understand the model’s reasoning process, allowing them to verify the output against their own domain knowledge and can boost the perceived helpfulness of LLMs. To initiate chain-of-thought reasoning, examples of chain-of-thought sequences are usually added to the prompt, guiding the LLM to apply this strategy in its generated output. \cite{Wei.2022}. 

Another common strategy is the use of \emph{personas}. Here, the LLM is asked to adopt the standpoint of a particular persona \cite{White.2023}, such as, for example, a journalist who needs to write an article. This strategy is particularly useful when users seek to write for a specific audience or from a specific standpoint as it allows the LLMs to decide what type of output it should generate and what details it should focus on \cite{White.2023}. As a result, using personas creates outputs that are often more personalized for different target audiences and can thus improve readers' perceptions. While there is evidence that these strategies influence the output quality of LLMs and users' experience \cite{Liu.2023, Wei.2022, White.2023}, little is known about the impact of \emph{training} novice users in prompt engineering. 

Previous works show that novice users often struggle with prompt engineering \cite{ZamfirescuPereira.2023, Jahani.2024, Dang.2022}. One reason for this is that they have an incomplete understanding of the capabilities of LLMs and they further tend to create prompts that mimic human-to-human instructions, for example, by relying more on instructions rather than giving examples \cite{ZamfirescuPereira.2023}. In addition, novice users tend to over-generalize from single observations and, hence, do not make systematic progress when engineering prompts \cite{ZamfirescuPereira.2023}. Other works have found that novice users interact with LLMs as they would with a human interlocutor, such as, for example, by using socially desirable phrases like \say{hello} and \say{thank you} and explaining inner thoughts and motives \cite{Knoth.2024}. Together, the above examples provide ample evidence that novice users do not know which information is important to be included in an effective prompt and, therefore, how to design effective prompts \cite{Knoth.2024}. This raises the need to understand how they can be trained in effective prompt engineering, which is the focus of our study. 

\subsection{Research gap}

Writing effective prompts is essential for leveraging the full potential of LLMs, but this can be particularly challenging for novice users. To the best of our knowledge, there is no work that analyzes the effect of training users in prompt engineering on user experience as well as output quality.  In this study, we are the first to train professional journalists in effective LLM prompting and assess the impact on user experience. (e.g., perceived expertise and perceived helpfulness) and output quality (e.g., accuracy as an objective metric but also reader perceptions such as clarity and engagement).

\section{Research questions}

In a preregistered study, we hypothesize and test whether training in prompt engineering can positively impact the user experience and output quality when interacting with LLMs. Our study specifically focuses on professional journalists as their tasks are representative of those performed by many other knowledge workers who process information in text form and generate reports. We thus study the following research questions (RQs):

\vspace{0.2cm}
\textbf{RQ 1}: \emph{How does prompt engineering training influence the user experience of journalists when interacting with LLMs?}
\vspace{0.2cm}

\noindent
We expect that, after training, users may find LLMs more helpful as they better understand how to generate the desired output through prompts. In addition, the training should also make users rate their expertise in using LLMs higher. As such, we define user experience along two dimensions: (i)~ the perceived expertise of journalists with the LLM and (ii)~ the perceived helpfulness of the LLM for a given task. To answer RQ1, we conduct a field experiment where we train $\emph{N}=29$ journalists in prompt engineering. In our field experiment, we task journalists with writing a short social media post (similar to those that are published on social media platforms such as Twitter/X and LinkedIn) with ChatGPT-3.5. This task is performed both before (henceforth called \emph{pre-training task}) and after (henceforth called \emph{post-training task}) receiving the training. As stipulated above, we measure the user experience -- i.e., perceived expertise and perceived helpfulness -- both before and after the journalists receive the training. 

Our field experiment is carefully designed to assess the effect of training in prompt engineering (see Sections~\ref{sec:field_experiment} and \ref{sec:RQ1} for details). In brevity, we randomize journalists into two groups that receive our tasks in opposite orders, where the tasks involve writing about two scientific articles, which we refer to as \textsc{Article Psy} and \textsc{Article GunViolence}. Put simply, one group receives \textsc{Article Psy} as a pre-training task and \textsc{Article GunViolence} as a post-training task (and vice versa for the other group). This way, we can isolate the effects of prompt engineering training on user experience.

\vspace{0.2cm}
\textbf{RQ 2}: \emph{How does prompt engineering training influence the accuracy of text written by journalists with LLMs?}
\vspace{0.2cm}

\noindent
In RQ2, we focus on accuracy as one measure of output quality, which is often relevant for domain experts. We expect that creating a social media post about a scientific article can potentially introduce inaccuracies in the text, either due to the more concise and engaging format of the social media post (as compared to the original scientific article) or due to LLM use because LLMs often tend to hallucinate \cite{Bang.2023}. Hence, we conduct an external, post-hoc evaluation of output quality, where we ask a domain expert to rate the accuracy of the texts that were written during the field experiment along a scoring system that penalizes inaccuracies (see Section~\ref{sec:rq2}). 

\vspace{0.2cm}
\textbf{RQ 3}: \emph{How does prompt engineering training influence the non-expert reader perception of texts written by journalists with LLMs?}
\vspace{0.2cm}

The generated social media posts may be perceived differently by domain experts (who may focus primarily on accuracy) and the general non-expert audience (who may prefer a social media post that is interesting and engaging). In RQ3, we thus examine the non-expert reader's perception of the written social media text as another dimension of output quality. Here, we consider how the general audience perceives the narrative, such as whether it effectively conveys the intended message and whether it can make them engage with the social media post. Hence, we conduct another external, post-hoc evaluation where we let non-expert readers evaluate the texts from the field experiment along several non-expert reader specific dimensions for assessing scientific journalism such as clarity and engagement (see Section~\ref{sec:rq3}).

\section{Field experiment}
\label{sec:field_experiment}

The experimental protocol was pre-registered at \url{https://osf.io/346dm/?view_only=d5f138787e944669911001a30d38ffee}.

\subsection{Overview}

We conducted a field experiment with a group of journalists specializing in science communication from Switzerland to measure the effect of prompt engineering training on user experience and output quality. The journalists ($N = 29$) participated in two writing tasks, namely, a pre-training task and a post-training task that took place before and after training, respectively. In both tasks, the journalists created a short social media post for an abstract of the scientific article using an LLM (specifically, ChatGPT-3.5). We used a counter-balanced measures design to examine the effect of prompt engineering training. Since treatment order can influence participants' behavior due to fatigue or other external factors \cite{Gaito.1961}, we randomly varied the order of the articles in the writing task. We also asked journalists to rate their experience with LLMs with respect to \emph{perceived expertise} and \emph{perceived helpfulness}.

The written social media posts subsequently went through two separate post-hoc external evaluations: (i)~an evaluation of accuracy by a domain expert and (ii)~an evaluation of reader perception (e.g., clarity and engagement) by a group of non-expert readers. We explain both post-hoc evaluations in Sections~\ref{sec:rq2} and \ref{sec:rq3}, respectively. 

\subsection{Tasks} 

In our field experiment, the participants (journalists) were asked to perform a science communication task. Participants should transform an extended abstract of a scientific article into an engaging Twitter/X post with the help of ChatGPT-3.5.\footnote{Participants used ChatGPT-3.5, accessed through the ChatGPT website interface on 25/04/2024.} We chose the task as it is common for science journalists in news-making \cite{TenenboimWeinblatt.2018}. We imposed a character limit of 280 characters for each post, similar to Twitter/X.  


For the writing tasks, we chose two scientific articles, which we refer to as \textsc{Article Psy} and \textsc{Article GunViolence}: 
\begin{itemize}
\item \textsc{Article Psy} (see \cite{Ross.2024}) was published in \emph{Journal of the American Medical Association}. It studies the effectiveness of psychiatric hospitalization on reducing subsequent suicidal behaviors in patients with suicidal ideation or suicide attempts and develops a machine learning model to tailor treatment recommendations based on individual patient characteristics \cite{Ross.2024}. 
\item \textsc{Article GunViolence} was published in \emph{JAMA Open} journal, which is also operated by the American Medical Association (see \cite{Semenza.2024}). It examines the association of exposure to gun violence with suicidal ideation and behaviors among black adults in the United States, thereby highlighting the significant impact of gun violence on mental health and the potential benefits of reducing interpersonal gun violence \cite{Semenza.2024}. 
\end{itemize}

\noindent
We provide the extended abstracts of the two scientific articles in Appendix~\ref{appendix:articles_abstract}. In our counter-balanced design, the participants conducted the writing tasks for both articles but with random article display orders; that is, half of the participants received \textsc{Article Psy} before training and \textsc{Article GunViolence} after, while the rest received the articles in the reversed order.

We intentionally selected the two articles from above as we expected several challenges that are common in science communication \cite{Jensen.2020, Guenther.2017}. This includes differentiating between correlation and causality (e.g., \textsc{Article GunViolence} offers associative but not causal findings), understanding the scope and boundary conditions of findings (e.g., \textsc{Article GunViolence} is limited to people identified as Black or African American in the US), interpreting data (e.g., statistical uncertainty), and translating jargon and technical language. Later, in RQ2, we also grade the accuracy of the social media posts along the previous dimensions (see Section~\ref{sec:rq2_method}).


\subsection{Training in prompt engineering}
\label{sec:prompt_engineering_training}

Our prompt engineering training for journalists was designed as a comprehensive 2-hour interactive training. All journalists attended the same session. This intensive session combined hands-on exercises with practical assignments, aiming to familiarise journalists with the capabilities of LLMs and train them in writing effective prompts. In developing this workshop, we drew upon best practices from both academic research \cite{Schulhoff.2024,Huang.2022,Mosbach.2023,Saravia.2022} and OpenAI's published guidelines for using ChatGPT \cite{OpenAI.2024}. For reasons of reproducibility, we make the slides of our course publicly available. We emphasize that several parts of our course involved interactive elements. The slides are available via a public repository: \href{https://github.com/vosh-96/The-Effect-of-Education-in-Prompt-Engineering-Evidence-from-Journalists}{https://github.com/vosh-96/The-Effect-of-Education-in-Prompt-Engineering-Evidence-from-Journalists}.

\begin{table}[htbp]
\footnotesize
\begin{tabular}{ll}
\toprule
\textbf{Topic} & \textbf{Key concepts} \\ 
\midrule
\multicolumn{2}{l}{\textbf{Part 1: Introduction to prompt engineering (55 minutes)}} \\
\midrule
Fundamentals of prompts & Importance and core elements of prompts \\
Prompt applications &  \multicolumn{1}{p{9cm}}{Text summarization, question and answer, text classification, and role playing}
 \\
Prompt  techniques & Zero shot, few shot, and chain-of-thought\\
\midrule
\multicolumn{2}{l}{\textbf{Part 2: Practical aspects (55 minutes)}} \\
\midrule
Prompt optimization & Iterative process for increasing the quality of prompts\\
Risks and challenges & Data bias, inequality, data privacy\\
\bottomrule
\end{tabular}%
\caption{Outline of the prompt engineering training.}
\label{tab:training_outline}
\end{table}

The primary objective of our training was to equip journalists with a robust understanding of prompt engineering principles and to develop their practical skills in crafting and refining prompts. As outlined in Table \ref{tab:training_outline}, the training was structured into two main parts. The first part laid the groundwork by introducing foundational concepts, beginning with an explanation of prompts and the concept of context length. We then progressed to explore various LLM applications, including text summarization, question answering, text classification, role-playing, and reasoning. This section concluded with an introduction to three key prompting strategies: zero shot, few shot, and chain-of-thought techniques. 
\begin{itemize}
    \item \emph{Zero shot}: The user provides a task description to the model without any further examples, and the model will generate the response based on the task description.
    \item \emph{Few shot}: Besides the task description, the user also provides a few example outputs in the desired format to the model. The model will use the task description and the examples to generate the response. 
    \item \emph{Chain-of-thought}: The user will ask the model to break down the task description into a series of intermediate steps and imitate human-like reasoning to solve them. 
\end{itemize}

The second part covered prompt refinement methods. Here, we focused on the iterative nature of prompting. We broke the process into three main subprocesses: \emph{getting started}, \emph{refining the focus}, and \emph{exploring in depth}. The getting started phase is mainly the first attempt to give the prompt context and objective. This was followed by iterating the prompt to improve the final result. We ended the training by reviewing the risks and ethical considerations of using LLMs in real life. 


Overall, participants showed enthusiasm for the technology. However, they had some serious reservations about the style of the output. While the outputs of LLM were generally regarded as being promising by the participants, the outputs still did not match the level of a seasoned journalist.

\subsection{Procedure}

\begin{figure}[htb]
\includegraphics[width=.9\linewidth]{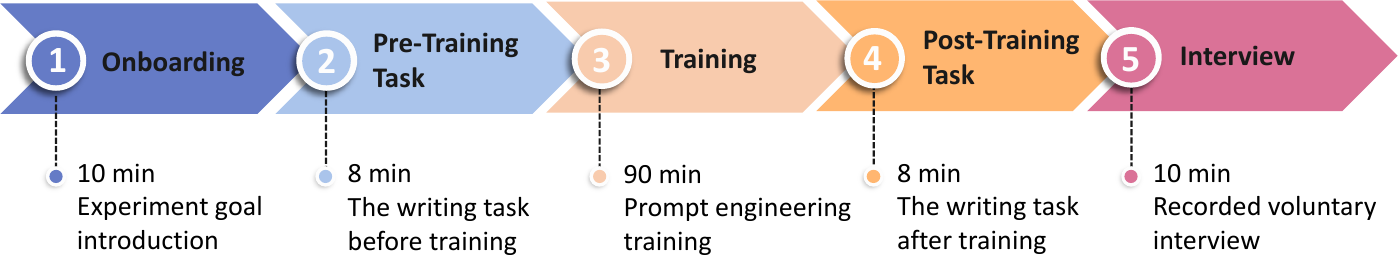}
\caption{Procedure of field experiment.}
\Description{Flowchart illustrating the sequence of steps in the field experiment, organized into five stages: Onboarding, Pre-Training Task, Training, Post-Training Task, and Interview. The Onboarding stage lasts 10 minutes and involves an introduction to the experiment. The Pre-Training Task stage takes 8 minutes and includes a writing task before training. The training stage is 90 minutes long and focuses on prompt engineering. The Post-Training Task stage involves an 8-minute writing task after training. The final stage, the Interview, lasts 10 minutes and includes a recorded voluntary interview. Each stage is annotated with its duration and a brief description of its content.}
\label{Fig:fieldexperiment}
\end{figure}

We conducted our field experiment on April 25, 2024, at the Swiss National Science Foundation (SNSF). Figure~\ref{Fig:fieldexperiment} shows the timeline of our field experiment. We started with onboarding, where we explained the goal of the experiment, briefly introduced prompt engineering, and collected informed consent. We then asked participants to fill in their demographic information, including age, gender, level of education, and years of experience, while the survey with demographic information was made optional. 

Next, participants were randomly assigned to two conditions, based on which we distributed the \emph{pre-training task}. Each group received either  \textsc{Article Psy} or \textsc{Article GunViolence} for writing social media posts. After completing the task, we asked them to rate (i)~the perceived expertise with ChatGPT and (ii)~the perceived helpfulness of ChatGPT in the task (both on a 7-point Likert scale).

Subsequently, we proceeded with the prompt engineering training (see Section~\ref{sec:prompt_engineering_training} for the training content). 

After training, we asked participants to perform the \emph{post-training task}, where the opposite task is now given to each group. This allows us to isolate the learning effect later. Again, we collected the participants' (i)~perceived expertise with ChatGPT and (ii)~perceived helpfulness of ChatGPT in the task. 

Lastly, the participants were given the opportunity to share their thoughts and experiences about using LLMs and prompt engineering for the task in qualitative form, and all participants were further invited to voluntary interviews. We debriefed the participants and appreciated their participation. 

\subsection{Participants}

We invited 37 journalists to our field experiment. They were non-incentivized but motivated due to the demand in their job to learn about prompt engineering for science communication. Participating and submitting the responses to the two writing tasks (\emph{pre-training task} and \emph{post-training task}), as well as demographics information, were optional. In the end, 29 participants provided valid responses.\footnote{We regard a response as valid if both writing tasks were completed.} We refer to the valid responses as \texttt{J1} through \texttt{J29}. One participant did not provide any demographics, but we still used the responses from the writing task to estimate the learning effect. Overall, the number of participants aligns with prior field experiments involving similar tasks for professionals \cite{Li.2024, Benk.2022, Osone.2021, Huang.2020, Senoner.2024}. 

Among the participants, 16 are women, 11 are men, and one is non-binary. The average age is 45.82 years, with an age range of 39 years (27 to 66). Except for one journalist with a Bachelor's degree, the rest of the participants held at least a Master's degree (or equivalent), and 14 even had a doctorate degree (or equivalent). Additionally, 16 participants had over 10 years of working experience.

\subsection{Ethical considerations}

We respect the privacy and autonomy of all participants involved in this study. Data were collected anonymously, with all personally identifiable information being removed. The study followed ethical research standards \cite{Rivers.2014, SalehzadehNiksirat.2023}, and the experimental design received approval from the Ethics Commission at the LMU Munich. We debriefed participants after the study and allowed them to ask questions and provide feedback. 


\section{Effect of training on user experience (RQ1)}
\label{sec:RQ1}

Our training in prompt engineering is aimed at helping journalists effectively use LLMs at their work. Hence, we expect successful training to improve the user experience with LLMs. Specifically, in RQ1, we are interested in knowing whether the training influenced the \emph{Perceived Expertise} and \emph{Perceived Helpfulness} of journalists when interacting with LLMs.

\subsection{Method}

In the field experiment, we asked the participants about their experience with LLMs before and after the prompt engineering training. Participants evaluated their (i)~perceived expertise and (ii)~perceived helpfulness of LLMs using a 7-point Likert scale ranging from 1 (``extremely unhelpful/no expertise'') to 7 (``extremely helpful/expert'').

As the collected data had a non-normal distribution, we used the Mann-Whitney U test \cite{McKnight.2010} to identify if there were statistically significant differences in the mean for (i)~perceived expertise and (ii)~perceived helpfulness when comparing the dimensions before and after training. In addition, we used two linear mixed-effects regression models with subject-level random effects to further quantify the role of control variables (e.g., demographics). The regression model is formalized as:
\begin{equation}
\small
\begin{aligned}
\emph{Y}_{i} &= \beta_0 + u_{0i} + \beta_1 \emph{TrainingStatus}_i + \beta_2 \emph{ArticleOrder}_i + \beta_3 \emph{Age}_i + \beta_4 \emph{Gender}_i + \beta_5 \emph{Education}_i + \beta_6 \emph{WorkExperience}_i + \epsilon_{i},
\end{aligned}
\label{eq:lrm}
\end{equation}
where $\emph{Y}_{i}$ denotes the experience with LLM (i.e., \emph{Perceived Experience} or \emph{Perceived Helpfulness}) for the $i$-th participant, with intercept $\beta_0$, subject-level random effects $u_{0i}$, coefficients $\beta_1$ to $\beta_6$, and an error term $\epsilon_{i}$. The other variables are as follows: \emph{TrainingStatus} represents the participant's training status, encoded as 0 (before training) and 1 (after training). \emph{ArticleOrder} indicates the order in which articles are displayed in the two tasks, with 0 representing \textsc{Article Psy} in the pre-training task and \textsc{Article GunViolence} in the post-training task, and 1 indicating the reverse order. \emph{Age}, \emph{Gender}, \emph{Education}, and \emph{WorkExperience} correspond to the participants' demographic variables. By examining the coefficient of the variable \emph{ArticleOrder}, we can assess the reliability of our counter-balanced study design, thus ensuring that the results are not influenced by the order in which the articles are displayed.

\subsection{Results}

\subsubsection{Perceived expertise in LLM use}

Figure~\ref{fig:perceivedexp} compares the perceived expertise before and after training. The participants' mean perceived expertise in LLM use increased by around 10\% from 3.38 to 3.72 after receiving prompt engineering training. Similarly, the median value increased from 3 to 4.


The regression results from Equation~\ref{eq:lrm} are reported in Table~\ref{tab:mixed_effect_lr}. We observed that \emph{TrainingStatus} had a positive and significant impact on perceived expertise ($\beta = 0.357$, $p < 0.05$). This finding confirms that prompt engineering training increases participants' perceived expertise. Specifically, all else being equal, training increased the perceived expertise score by approximately 0.357 units, highlighting the effectiveness of the prompt engineering training in enhancing participants' self-assessed capabilities in leveraging LLMs. Conversely, \emph{Age} had a significant negative effect on perceived expertise ($\beta = -0.083$, $p < 0.05$), while other demographic variables had $p$-values above 0.05. This result suggests that, if holding other factors constant, older participants perceive themselves as less proficient with LLMs. Furthermore, the regression analysis showed that \emph{ArticleOrder} was not significant ($\beta = 0.063$, $p > 0.05$). This confirmed that the order in which the articles were presented to participants did not influence their perceived expertise, thus confirming the reliability of our study design.


\subsubsection{Perceived helpfulness of LLM use}

\begin{figure}[h]
  \centering
  \begin{minipage}{0.4\textwidth}
    \includegraphics[width=1.1\linewidth]{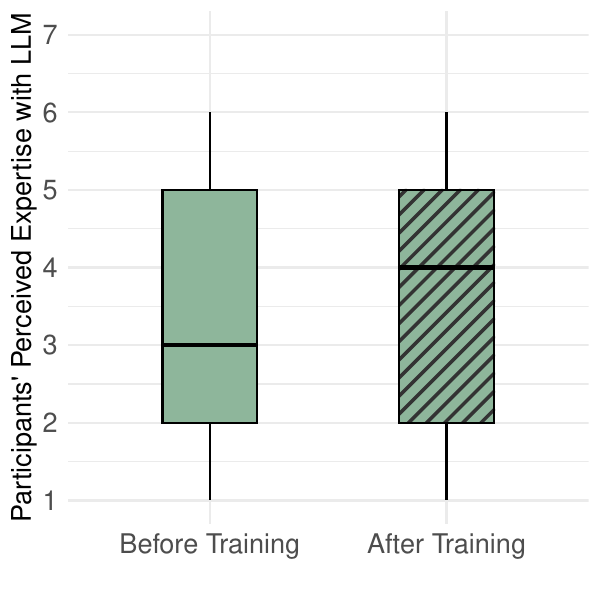}
    \caption{Participants' perceived expertise in using LLMs before and after training (\emph{N=}29).}
    \Description{Boxplots of participants’ perceived expertise in using large language models before and after training, measured on a scale of 1 to 7. Left boxplot represents perceived expertise before training, with an interquartile range spanning approximately 2 to 5 and a median value of 3. Right boxplot represents perceived expertise after training, with an interquartile from 2 to 5 and a median of 4. The sample size is 29 participants (N=29)}
    \label{fig:perceivedexp}
  \end{minipage}\hfill
  \begin{minipage}{0.4\textwidth}
    \includegraphics[width=1.1\linewidth]{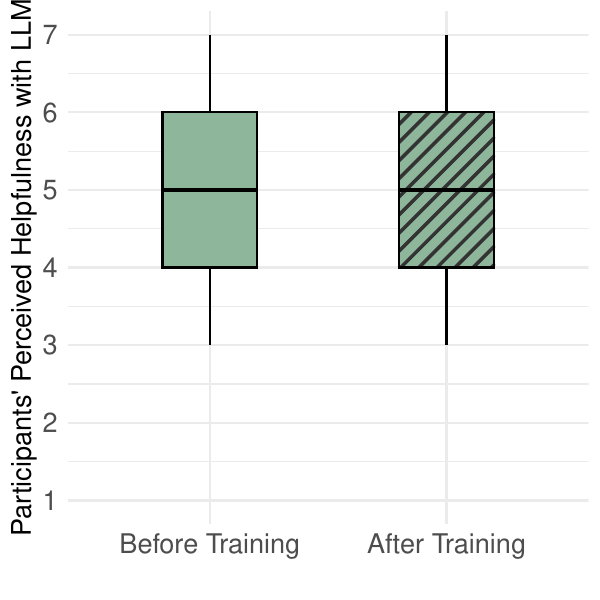}
    \caption{Participants' perceived helpfulness of LLMs before and after training (\emph{N=}29).}
    \Description{Boxplots of participants’ perceived helpfulness of large language models before and after training, measured on a scale from 1 to 7. The left boxplot represents perceived helpfulness before training, with an interquartile range from approximately 4 to 6 and a median value of around 5. The right boxplot represents perceived helpfulness after training, showing a similar interquartile from 4 to 6, with a median around 5. Both conditions display a consistent spread of responses, with the middle 50\% of participants rating the helpfulness within the same range before and after training. The sample size is 29 participants (N=29).}
    \label{fig:perceivedhelp}
  \end{minipage}
\end{figure}

Our study finds that, on average, the journalists' perceived helpfulness of LLMs declines after training. The mean value of perceived helpfulness decreased by 3.5\% from 4.93 to 4.76, while the median value remained at 5 (see Figure~\ref{fig:perceivedhelp}). 

Next, we estimated the effect of the prompt engineering training, demographics, and article display order on participants' perceived helpfulness based on Equation~\ref{eq:lrm}. The regression results are in Table~\ref{tab:mixed_effect_lr}. We observed that \emph{TrainingStatus} even had a negative coefficient ($\beta = -0.286$, $p = 0.073$) but was slightly above common significance thresholds (potentially due to our sample size).  This may still suggest that receiving prompt engineering training could even reduce the journalists' perceived helpfulness with LLMs. All demographic variables had $p$-values above 0.05. Furthermore, \emph{ArticleOrder} was again not statistically significant ($\beta = -0.241$, $p > 0.05$), which confirms the reliability of the study design. 

During our oral interviews with the journalists, we obtained more insights regarding why the perceived helpfulness decreases. The journalists acknowledged the efficiency of ChatGPT in sorting information but criticized its stylistic quality, commenting it was similar to that of a ``secondary school dissertation''. Although the tool was able to significantly speed up the writing process, the final output often required substantial revisions to meet the journalists' standards. Some journalists got frustrated when ChatGPT did not follow given commands, which diminished their creativity and led to a reluctance to use it consistently. The participants also raised concerns about ChatGPT's cultural adaptability. In particular, the journalists in our field experiment mentioned that tools developed in the US might not be well-suited to European cultural contexts. We provide further details about the oral interviews in Appendix~\ref{appendix:oral_interview_records}.

\begin{table}[h]
\centering
\small
\begin{tabular}{lccccccc}
\toprule
    & \emph{ArticleOrder} & \emph{TrainingStatus} & \emph{Age} & \emph{Gender} & \emph{Education} & \emph{WorkExperience} & \emph{Intercept}  \\
\midrule                      
\begin{tabular}[c]{@{}l@{}}\emph{Perceived Expertise}\end{tabular}   
& \begin{tabular}[c]{@{}c@{}}0.063\\ (0.551)\end{tabular}    
& \begin{tabular}[c]{@{}c@{}}0.357*\\ (0.164)\end{tabular}
& \begin{tabular}[c]{@{}c@{}}$-$0.083*\\ (0.033)\end{tabular}
& \begin{tabular}[c]{@{}c@{}}0.800\\ (0.624)\end{tabular}
& \begin{tabular}[c]{@{}c@{}}0.026\\ (0.478)\end{tabular}
& \begin{tabular}[c]{@{}c@{}}0.217\\ (0.215)\end{tabular}
& \begin{tabular}[c]{@{}c@{}}6.228***\\ (1.556)\end{tabular}\\

\begin{tabular}[c]{@{}l@{}}\emph{Perceived Helpfulness}\end{tabular}   
& \begin{tabular}[c]{@{}c@{}}$-$0.241\\ (0.437)\end{tabular}    
& \begin{tabular}[c]{@{}c@{}}$-$0.286$^.$\\ (0.153)\end{tabular}
& \begin{tabular}[c]{@{}c@{}}$-$0.037\\ (0.026)\end{tabular}
& \begin{tabular}[c]{@{}c@{}}0.726\\ (0.495)\end{tabular}
& \begin{tabular}[c]{@{}c@{}}$-$0.212\\ (0.379)\end{tabular}
& \begin{tabular}[c]{@{}c@{}}$-$0.043\\ (0.170)\end{tabular}
& \begin{tabular}[c]{@{}c@{}}6.965***\\ (1.235)\end{tabular}\\
\bottomrule  
\end{tabular} \\
Significance levels: $^.p<0.1$, $^*p<0.05$, $^{**}p<0.01$, $^{***}p<0.001$
\caption{Mixed-effect linear regression results.}
\label{tab:mixed_effect_lr}
\end{table}

\subsection{Interpretation of findings}


In RQ1, we test the effect of prompt engineering training on journalists' (i) perceived expertise and (ii) perceived helpfulness when interacting with LLMs. We find that our training increased the journalists' expertise in using ChatGPT. This suggests that structured training in prompt engineering can empower users by enhancing their understanding of how to interact with LLMs effectively. This finding aligns with previous research indicating that AI literacy can be promoted through targeted training (e.g., \cite{Kandlhofer.2016, Markus.2024}). Notably, older populations rate their expertise lower than younger populations, which points to a potential barrier when adopting LLMs and which may require targeted interventions to enhance their confidence and perceived skill levels.

However, an improvement in perceived expertise does not necessarily translate into an improvement in perceived helpfulness of using LLMs. Our results may even imply that training leads to a decrease in perceived helpfulness, but this could be attributed to the unmet expectations of the journalists in our experiment. One possible reason is that end-users do not fully understand the inherent limitations of LLMs, such as hallucinations and reasoning constraints and have overly positive perceptions about them initially. Our findings align well with previous work in that AI literacy training can help put the abilities of AI tools into perspective and make limitations more apparent \cite{Theophilou.2023}. This, in turn, is likely to lower the perceived helpfulness of such tools. Another reason could be that most of our participants are professional journalists with over ten years of experience and at least a Master's degree. They are likely to have high standards for creativity and insight in their creative work. Such a group of experienced professionals may find AI-generated content to lack originality and sophistication; hence, LLMs may be perceived as less helpful in comparison to their own capabilities. This aligns with findings from previous studies that experienced professionals often see less value in AI-generated content when it does not match their quality standards \cite{Li.2024}.

\section{Effect of training on accuracy (RQ2)}
\label{sec:rq2}

Next, we assess the accuracy of the posts that the participants wrote. This is particularly important due to the challenging nature of the task, which involves distinguishing correlational and causal language and interpreting field-specific jargon. In addition, LLMs tend to hallucinate and produce wrong output, which needs to be accounted for. 

\subsection{Method}
\label{sec:rq2_method}

To assess the \textbf{accuracy} of the output, we enlisted a domain expert, in this case, a professor with experience in both machine learning and medical research at a renowned European university, to conduct a blind evaluation. To ensure reproducibility, we developed a robust scoring system that awards or penalizes each post by one point based on specific criteria (Table~\ref{tab:expert_score_criteria}). Our scoring system distinguishes different dimensions: (1)~We refer to \emph{factual errors} when there is a misuse of casual language or incorrect information. We penalize a post with a minus point once for each factual error (e.g., misuse of causal language). (2)~We refer to \emph{representation errors} as ones that convey a misleading or incorrect narrative as compared to the original materials. Here, we penalize each representation error again with one minus point (e.g., these are mostly due to hallucination of the LLM, such as using hashtags that are unrelated to the research). (3)~We reward posts for having depth (e.g., mentioning limitations of the study). Finally, we calculate an \emph{overall score} by summing over the points awarded for depth minus the negative points for factual errors and representation errors. Table~\ref{tab:tweet_example} gives two examples of participant-generated posts. For each example, we provide feedback from the domain expert about why the post lacks accuracy. 

Given the non-normal distribution of the collected data, we employ Yuen's test \cite{Wilcox.2011} for trimmed means to conduct between-subject comparisons of each article in terms of the above accuracy scores before and after training. To account for the nested structure of the data, we implement a linear mixed-effects regression model with a random intercept for articles. This approach allows for considering within-article variability while controlling for potential confounding effects associated with specific articles. The regression model is given by:
\begin{equation}
\small
\begin{aligned}
\emph{S}_i &= \beta_0 + u_{0i} + \beta_1 \emph{TrainingStatus}_i +  \beta_2 \emph{Age}_i + \beta_3 \emph{Gender}_i + \beta_4 \emph{Education}_i + \beta_5 \emph{WorkExperience}_i + \epsilon_{i},
\end{aligned}
\label{eq:lrm_expert}
\end{equation}
where $\emph{S}_{i}$ is the accuracy based on our scoring system (i.e., factual errors, representation errors, depth, and the overall score). For better interpretability, the overall score for the $i$-th participant is normalized to the range [0, 1] at the article level. The above regressions model includes a fixed intercept $\beta_0$, article-level random effects $u_{0i}$, fixed effects coefficients $\beta_1$ to $\beta_5$, and an error term $\epsilon_i$.

\begin{table}[htbp]
\resizebox{\textwidth}{!}{%
\begin{tabular}{lll}
\toprule
\textbf{Factual errors} (subtract 1 point)    & \textbf{Representation errors} (subtract 1 point)                                                  & \textbf{Depth} (receive 1 point)                        \\ \midrule
\multicolumn{2}{l}{\textsc{Article Psy}}                                                                                                        \\ \midrule
Misuse of casual language                 & Lack of effect size mentioned & Specify subjects being veterans
                                                   \\
Misuse of terminologies     &  Insufficient consideration of heterogeneity effects on study outcomes & Highlight limitations    \\
Incorrect report of numbers  &  Write the post as they are the researchers
& Mention of time-frame of outcome or of population              \\
Incorrect report of concepts & Unrelated hashtags
                         & Complete mention of outcomes                    \\
 & Incorrect mention of the source of heterogeneity                                            & Mention of the study's baseline as current practice \\ \midrule
\multicolumn{2}{l}{\textsc{Article GunViolence}}                                                                                                        \\ \midrule
Misuse of casual language   & Absence of reported effect sizes and quantitative results &                                    Specifying geographical context (e.g., US, American)  \\
Incorrect report of study's subject & Incorrect policy implications  & Highlight limitations                         \\
Incorrect numbers in the posts & Write the post as they are the researchers & Mention of both outcomes \\
Incorrect massages & Mix different results & Mention detailed inclusion criteria \\ \bottomrule
\end{tabular}%
}
\caption{Scoring criteria for evaluating the accuracy of posts (through a domain expert). The base score is zero.}
\label{tab:expert_score_criteria}
\end{table}

\begin{table}[htbp]
\resizebox{\textwidth}{!}{%
\begin{tabular}{ll}
\toprule
\textsc{Article Psy}  & \textsc{Article GunViolence}   \\ \midrule
  \multicolumn{1}{p{10cm}}{\textbf{Post:} \emph{``New study shows psychiatric hospitalization reduces immediate suicide attempt risk but varies by patient history. Precision treatment could cut suicide attempts by 16\% and reduce hospitalizations by 13\%. \#MentalHealthAwareness \#SuicidePrevention``}}    & \multicolumn{1}{p{10cm}}{\textbf{Post:} \emph{``Hey, young changemakers! Did you know gun violence hits Black youth hard, leading to higher rates of suicidal ideation \& attempts? Let's unite to demand safer communities for our generation. \#YouthForChange \#MentalHealthAwareness``}}    \\ \midrule 
\multicolumn{1}{p{10cm}}{\textbf{Accuracy assessment:} The use of {``reduce''} suggests casual language (\emph{factual error}). The post further fails to address the study's heterogeneity in its sample (representation error). However, it successfully covers multiple research outcomes (\emph{depth}).} &  \multicolumn{1}{p{10cm}}{\textbf{Accuracy assessment:} The post implies casual language and incorrectly reports the subjects as {``youth''} (2x \emph{factual error}). The post does not mention any quantitative result or effect size (\emph{representation error}). The post mentions both outcomes (\emph{depth}).} \\ \bottomrule
\end{tabular}%
}
\caption{Examples of written Twitter/X posts with additional feedback about the accuracy from the domain expert.}
\label{tab:tweet_example}
\end{table}

\subsection{Results}

\begin{figure}[h]
    \includegraphics[width=0.8\linewidth]{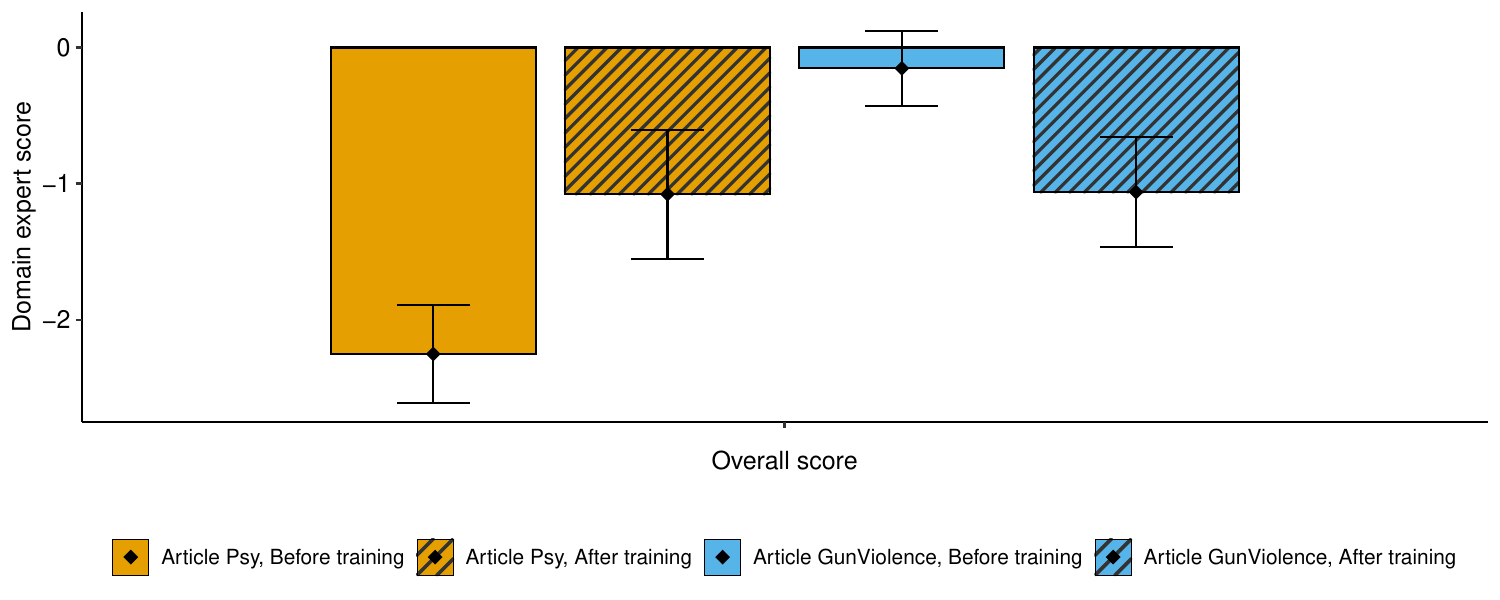}
    \caption{The overall score measuring accuracy (as assessed by the domain expert). Whiskers refer to standard errors.}
    \Description{The domain expert overall score of the post for each article pre and post training. The overall score increased for Article Psy and decreased for Article GunViolence. }
    \label{fig:expert_score_overall}
\end{figure}

\begin{figure}[h]
  
    \includegraphics[width=0.95\linewidth]{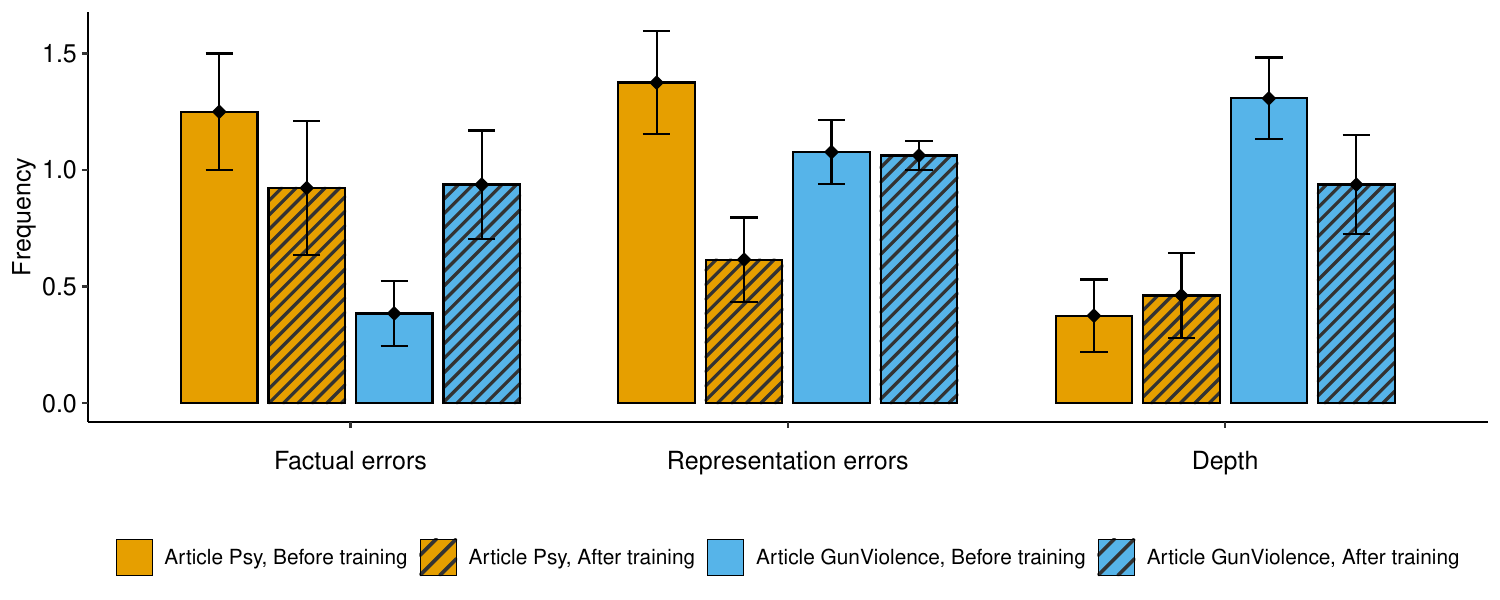}
    \caption{The average number of factual errors, misrepresentation errors, and bonus points for depth. All scores were evaluated by the domain expert. Whiskers refer to standard errors.}
    \Description{The figure shows the bar charts with standard errors for both articles before and after training. The y axis the frequency of expert metric existence in a post. The x axis three expert scores: factual errors, representation errors and depth. For Article Pys, the number of factual and representation errors decreased after training, and the reward for depth remained almost the same. However, for Article Gun Violence, factual errors increased, the reward for the depth of the post decreased, and the representation errors stayed the same after training.}
    \label{fig:expert_score_detailed}

\end{figure}

The comparison of the overall score is in Figure~\ref{fig:expert_score_overall}. A detailed comparison of factual errors, representation errors, and depth is in Figure~\ref{fig:expert_score_detailed}.

\textbf{Factual errors} (Figure~\ref{fig:expert_score_detailed}, left): For \textsc{Article Psy}, the mean number of factual errors per post reduced from 1.33 ($95\%$ CI = [0.81, 1.85]) before training to 0.92 ($95\%$ CI = [0.30, 1.55]) after training. Conversely, it increased for \textsc{Article GunViolence} from 0.38 ($95\%$ CI = [0.08, 0.69]) before training to 0.94 ($95\%$ CI = [0.44, 1.43]) after training. This trend aligns with the overall score from above, where \textsc{Article Psy} showed improvement with regard to the mean overall score from $-$2.31 ($95\%$ CI = [$-$3.09, $-$1.54]) to $-$1.08 ($95\%$ CI = [$-$2.11, $-$0.05]) post-training, while, for \textsc{Article GunViolence}, the mean overall score decreased from $-$0.15 (CI = [$-$0.75, 0.44]) to $-$1.06 (CI = [$-$1.92, $-$0.20]) after training. However, caution is warranted as the results are not statistically significant at common significant thresholds.

\textbf{Representation errors} (Figure~\ref{fig:expert_score_detailed}, center): The mean number of representation errors decreased for \textsc{Article Psy} from 1.38 ($95\%$ CI = [0.90, 1.85]) to 0.62 ($95\%$ CI = [0.22, 1.01]). For \textsc{Article GunViolence}, it remained relatively unchanged; i.e., it amounted to 1.08 ($95\%$ CI = [0.78, 1.38]) before training and 1.06 ($95\%$ CI = [0.93, 1.20]) after training. 


\textbf{Depth} (Figure~\ref{fig:expert_score_detailed}, right): The mean score for depth per post increased for \textsc{Article Psy} from 0.33 ($95\%$ CI = [0.01, 0.66]) to 0.46 ($95\%$ CI = [0.06, 0.86]) after training. In contrast, it decreased for \textsc{Article GunViolence} from 1.31 ($95\%$ CI = [0.93, 1.69]) to 0.94 ($95\%$ CI = [0.48, 1.39]) after training. 


\subsection{Interpretation of findings}

For RQ2, we find that the accuracy increases for \textsc{Article Psy} and decreases for \textsc{Article GunViolence}. Evidently, the divergent outcomes are consistent across all three measures (i.e., factual errors, representation errors, and depth), which indicates that the accuracy may be associated with the article's complexity. In other words, this suggests that the effectiveness of prompt engineering training in terms of accuracy may be task-dependent. Hence, the complexity of the content appears to influence the extent to which journalists can apply their training effectively. This finding is in line with research that indicates task complexity can impact the efficacy of AI tools more broadly \cite{Salimzadeh.2023}, while we add new evidence for prompt engineering.

\section{Effect of training on reader perception (RQ3)}
\label{sec:rq3}
Next, we assess the \textbf{reader perceptions} of the Twitter/X posts across a general audience that was initially targeted by the journalists (i.e., average readers of news). To do so, we conducted a pre-registered, post-hoc evaluation using non-experts recruited from \url{prolific.com} to evaluate the quality of posts created by our journalists both before and after training.

\subsection{Method}
\label{sec:rq3_method}

\begin{figure}[htb]
\includegraphics[width=1\linewidth]{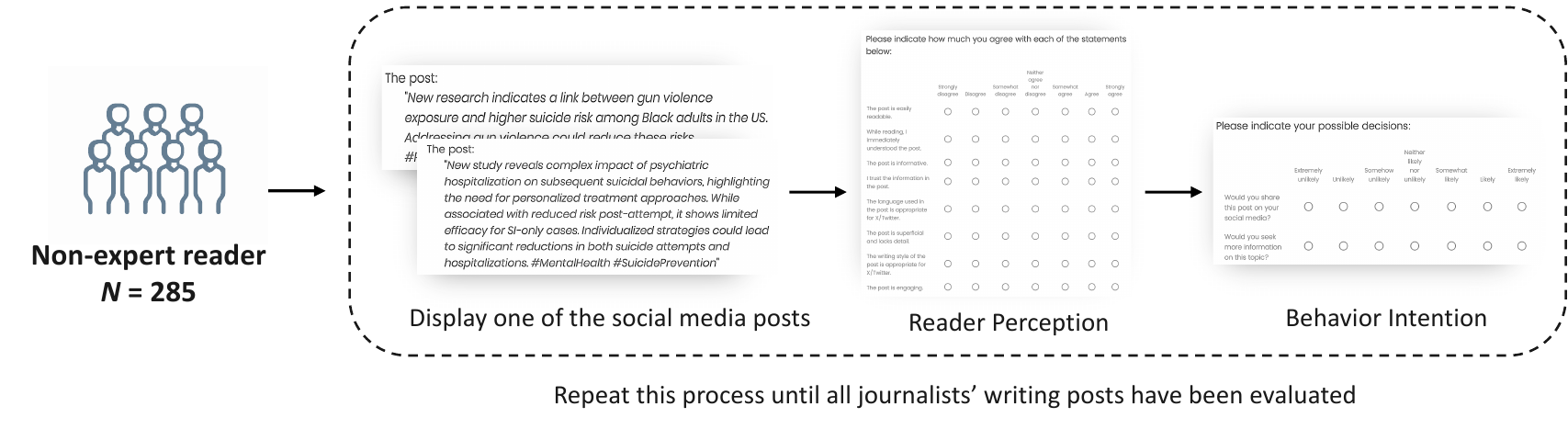}
\caption{Procedure for external, post-hoc evaluation in terms of non-expert reader perceptions (e.g., clarity, engagement).}
\Description{Flowchart depicting the evaluation process for journalists’ writings by non-experts. The process begins with recruiting a group of non-expert evaluators, represented by an icon on the left. They are shown a writing excerpt, followed by two forms: one for assessing reader perception and another for behavior intention. The forms consist of Likert scales for different statements. The flow is linear, moving from left to right, and the process repeats until all journalists’ writings have been evaluated. The diagram includes dashed boxes around the forms to indicate the steps involved.}
\label{Fig:externalevaluation}
\end{figure}
An overview of the post-hoc evaluation is shown in Figure~\ref{Fig:externalevaluation}. We examine the effect of prompt engineering training on ten different dimensions of text quality and reader response, as evaluated by non-expert readers. These dimensions are:
(1)~readability, (2)~clarity, (3)~informativeness, (4)~perceived trustworthiness, (5)~appropriateness for the target audience, (6)~depth, (7)~appropriateness for the mode of communication, (8)~engagement, (9)~intention to re-share, and (10)~intention to seek further information.
A full overview is in Table~\ref{tab:post_hoc2_var}.


The above dimensions have been widely employed in the literature (e.g., \cite{vanderLee.2019,Celikyilmaz.2020}) to assess the quality of text from a reader's perspective. Our hypothesis is that prompt engineering training may positively influence these aspects of text quality and reader response. The first eight dimensions focus on intrinsic text qualities, while the last two measure behavioral intentions such as information seeking and information sharing, which are both important factors for news consumption and social media more broadly \cite{Prollochs.2023}. Together, these dimensions provide a comprehensive evaluation of both content characteristics and potential reader actions. We used Yuen’s test \cite{Wilcox.2011} for trimmed means to compare the dimensions before/after training.

\begin{table}[htbp] 
\footnotesize
\begin{tabular}{ll} 
\toprule
  \textbf{Variable name} & \textbf{Question}  \\ 
\midrule 
 \multicolumn{1}{p{4.5cm}}
{\textbf{Intrinsic text quality} \newline \emph{ Readability} \newline \emph{Clarity} \newline \emph{Informativeness}  \newline \emph{Perceived trustworthiness}  \newline \emph{Appropriateness for the target audience}  \newline \emph{Depth} \newline \emph{Appropriateness for the mode of communication} \newline \emph{Engagement}} & \multicolumn{1}{p{6cm}}{~ \newline \emph{The post is easily readable.} \newline \emph{While reading, I immediately understood the post.} \newline \emph{The post is informative.} \newline \emph{I trust the information in the post.} \newline \emph{The language used in the post is appropriate for X/Twitter.} \newline \emph{The post is superficial and lacks detail.} \newline \emph{The writing style of the post is appropriate for X/Twitter.} \newline \emph{The post is engaging.}}\\
\midrule
\multicolumn{1}{p{4.5cm}}{\textbf{Behavioral intentions} \newline \emph{Intention to re-share} \newline \emph{Intention to  seek further information }} & \multicolumn{1}{p{6cm}}{~ \newline \emph{Would you share this post on your social media?} \newline \emph{Would you seek more information on this topic?}}\\
\bottomrule
\end{tabular}
\caption{Summary of variables used for assessing the reader perception and the corresponding survey questions. All variables were collected on a 7-point Likert scale.} 
\label{tab:post_hoc2_var}
\end{table}

\subsubsection{Procedure}

Prior to participation, we obtained informed consent from all non-expert readers. Subsequently, participants completed an attention check. It is noteworthy that we retained all participants in our primary analysis, regardless of their performance on the attention check. However, we conducted a robustness check excluding those who failed the attention check, which yielded consistent results. We randomly assigned each participant to a pair of social media posts from the same journalist so that we could later rule out between-journalist variability in terms of style. For each post, participants responded to the above-mentioned questions along 10 dimensions (i.e., eight pertaining to intrinsic text quality and two pertaining to behavioral intentions). All responses were recorded on a 7-point Likert scale. To mitigate order effects, the sequence of X/Twitter posts was randomized for each participant in the post-hoc evaluation. 

\subsubsection{Participants}

We initially collected responses from 318 participants through the Prolific platform. To maintain data integrity and minimize confounding factors, we implemented strict inclusion criteria. Specifically, we excluded participants who failed to provide their unique Prolific ID in the survey or those who attempted to complete the survey multiple times. This exclusion process was crucial to ensure that each non-expert reader was exposed to posts from only one journalist, thereby eliminating potential bias in our data. The final number of accepted non-expert readers is 285; on average, each pair of X/Twitter posts created by the journalists received 19.65 evaluations. Their ages range from 19 to 77 years old, with a mean of 36.13 years and a median of 33 years. Out of the 285 non-expert readers, 124 are men, 155 are women, 4 identified as non-binary, and 2 preferred not to answer.

\subsection{Results}

\begin{figure}[htb]
\includegraphics[width=1\linewidth]{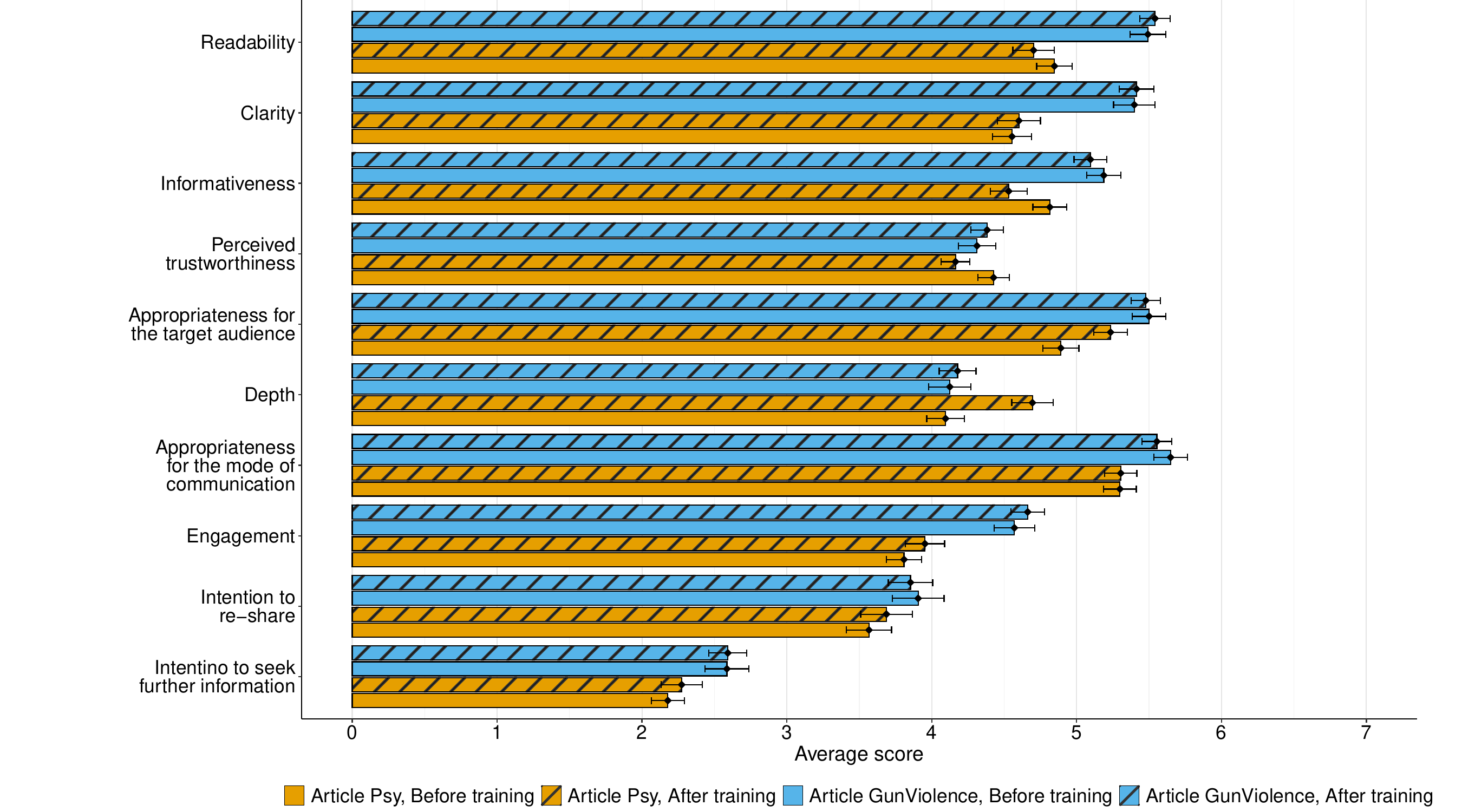}
\caption{External evaluation of reader perception (assessed by non-expert readers). The bars represent the score averaged over all non-expert readers and posts from the corresponding article. Whiskers represent standard errors.}
\Description{External evaluation of reader perception (assessed by non-expert readers). The bars represent the score averaged over all non-expert readers and posts from the corresponding article. The graph shows ten metrics which eight of them represent intrinsic text quality, and two represent the behavioral intentions. The effects of training on \textsc{Article Psy} and \textsc{Article GunViolence} varied across different metrics.}
\label{fig:externalevaluation}
\end{figure}

The results are shown in Figure~\ref{fig:externalevaluation}. The effects of training on \textsc{Article Psy} and \textsc{Article GunViolence} varied across different metrics. For \textsc{Article Psy}, improvements were in \emph{Appropriateness for the target audience}, which increased from 4.89 ($95\%$ CI: [4.65, 5.14]) to 5.23 ($95\%$ CI: [5.01, 5.46]), and \emph{Depth}, which increased from 4.10 ($95\%$ CI: [3.84, 4.35]) to 4.70 ($95\%$ CI: [4.41, 4.98]). For \textsc{Article GunViolence}, improvements were seen in \emph{Engagement}, which increased from 4.57 ($95\%$ CI: [4.29, 4.85]) to 4.66 ($95\%$ CI: [4.43, 4.89]). Interestingly, some metrics decreased slightly after training, such as \emph{Informativeness} for both articles. For example, for \textsc{Article Psy}, the \emph{Informativeness} dropped from 4.82 ($95\%$ CI: [4.58, 5.05]) to 4.53 ($95\%$ CI: [4.28, 4.78]), while, for \textsc{Article GunViolence}, it decreased from 5.19 ($95\%$ CI: [4.95, 5.42]) to 5.10 ($95\%$ CI: [4.87, 5.32]). The \emph{Intention to seek further information} remained relatively stable for both articles, with \textsc{Article Psy} showing a minimal increase from 2.18 ($95\%$ CI: [1.95, 2.40]) to 2.27 ($95\%$ CI: [1.99, 2.55]), and \textsc{Article GunViolence} maintaining the same mean of 2.59 with slightly adjusted confidence intervals. However, for none of the variables, the difference is statistically significant at common thresholds.

\subsection{Interpretation of findings}

In RQ3, we tested the effect of training on reader perception across text quality and behavioral intentions. Our findings indicate that prompt engineering training had a varied impact on the perceived quality of X/Twitter posts by non-expert readers. Specifically, there may have been slight improvements in dimensions like \emph{Style Appropriateness} and \emph{Depth}, suggesting that journalists may have been able to craft more tailored and detailed content after receiving structured guidance. This aligns with the idea that prompt engineering strategies such as chain-of-thought reasoning and persona adoption help to improve the relevance of output text for the target audience \cite{Wei.2022, White.2023}.

Further, the small changes in \emph{Engagement} and the slight decline in \emph{Informativeness} may suggest that prompt engineering training can improve technical aspects of writing but does not necessarily translate into more compelling or informative posts from a reader's perspective. Moreover, the \emph{behavioral intentions} of readers to seek further information or to share the posts are fairly similar, which could imply that the training did not significantly impact the overall appeal or effectiveness of the posts in driving reader action. This could be due to the complex nature of social media engagement, where clarity and engagement might be more critical than the technical accuracy or depth that prompt engineering tends to enhance.

However, we are careful with our interpretation as the above comparisons were not statistically significant. This may indicate that LLMs can generate content that is of a similar style as professional journalists and thus reach a high level of text quality in the eyes of non-expert readers. Hence, our findings suggest that it may be difficult for non-expert readers to identify meaningful differences between content that is LLM-generated or human-generated. Similar observations were made in studies comparing how people perceive misinformation generated by LLMs vs. humans, finding that non-experts cannot distinguish the veracity of both \cite{Bashardoust.2024}.  

The above analysis underscores that prompt engineering training should not only focus on the efficiency gains for LLM users but should also account for the downstream impact on the perception of non-expert readers. Hence, a good prompt engineering strategy is one that not only improves the precision and depth of content but also addresses how to maintain -- or even enhance -- information-seeking and reader engagement. This is particularly important in the context of social media, where behavioral intentions to reshare content are crucial for social media posts to go viral. Our findings align with the broader literature, suggesting that, while AI literacy and prompt engineering are crucial for leveraging LLMs effectively, they must be adapted to the specific needs of the task and audience to be truly impactful in professional settings such as journalism \cite{ZamfirescuPereira.2023, Ng.2021}.

\section{Further analyses}

\textbf{Text similarity:} Previous research has suggested that generative AI enhances individual creativity but reduces the collective diversity of novel content \cite{Doshi.2023}. Our study extends this analysis by evaluating the diversity of journalists' written posts after receiving prompt engineering training. 

We first cleaned the data for all written posts and then used document embeddings from Huggingface \cite{Mikolov.2013, Google.2013}, which were pre-trained using Google News and should match the style of journalistic writing. Next, we subdivided the posts into four subgroups: (1)~Before Training - \textsc{Article Psy}, (2)~Before Training - \textsc{Article GunViolence}, (3)~After Training - \textsc{Article Psy}, and (4)~After Training - \textsc{Article GunViolence}. We applied cosine similarity \cite{Rahutomo.2012} to calculate the average similarity scores for all posts in each subgroup (ranging from 0 to 1). To examine changes in diversity, we used the Mann-Whitney U test \cite{McKnight.2010, Nachar.2008}to compare the similarity scores before and after training but separately for textsc{Article Psy} and \textsc{Article GunViolence}  (for example, we compare the similarity scores from Before Training - \textsc{Article Psy} vs. After Training - \textsc{Article Psy}). A significant increase in the average similarity score indicates that the posts have become more similar to each other, suggesting a decrease in diversity. Conversely, a decrease in the average similarity score would indicate increased diversity among the posts.

For \textsc{Article Psy}, the average similarity score of the written posts drops significantly after training from 0.829 to 0.789 ($p < 0.05$), indicating increasing diversity in the content. Conversely, for \textsc{Article GunViolence}, the average similarity score increases from 0.827 to 0.857 ($p < 0.05$), suggesting that the content becomes more homogeneous. The mixed results suggest that the effectiveness of prompt engineering training in enhancing content diversity may vary depending on the topic and the participant's engagement with the material.

\textbf{Interpretation of findings:} Prompt engineering training can help journalists create more diverse content with ChatGPT if they effectively utilize the prompting strategies and refinement methods. However, it is important to recognize that diversity is not the sole criterion for evaluating output content; the task topic or inherent randomness in ChatGPT's outputs may also influence diversity. Overall, our similarity analysis points to a positive impact of prompt engineering training on the diversity of journalists' output content. 

\textbf{Manual analysis of the chat dialogues in ChatGPT:} We manually analyzed the participants' chat dialogues with ChatGPT. We observed interesting patterns. In instances where the similarity score is lower than the average for the same subgroup, journalists tend to have more substantive interactions and inputs with ChatGPT.\footnote{Here, we excluded prompts like \texttt{``Do you speak French?''} or \texttt{``Can you draw a picture?''} from our manual analysis.}. For example, in the pre-training task for \textsc{Article Psy}, one participant (\texttt{J11}) interacted with ChatGPT 12 times and thus obtained the lowest score amount to 0.727 (the subgroup average is 0.829), pointing to a post that is more diverse than the rest. Participant \textsc{J11} had also the lowest similarity score (0.603, while the subgroup average was 0.789) in the post-training task for \textsc{Article Psy}, against pointing to a post that is highly diverse from the rest. We found that the inputs of participant \texttt{J11} were generally effective since the participant clearly asked ChatGPT to craft or revise the posts. Similarly, \texttt{J26}, who obtained a similarity score of 0.749 (subgroup average is 0.827), interacted with ChatGPT six times in the pre-training task on \textsc{Article GunViolence}. The input of participant \texttt{J26} was also clear and directive, such as \texttt{``add the proportion of black people''} or \texttt{``add the publication source''}. In contrast, most participants who scored high in similarity and thus generated less creative content tended to use fewer effective prompts. For example, they often simply pasted the article and prompted with something simple such as \texttt{``Write a tweet with 280 characters''}.

These patterns point to the influence of our prompt engineering training, which introduces prompting strategies and refinement methods (see Section~\ref{sec:prompt_engineering_training}). In the case of participant \textsc{J11}, after training, the journalist added detailed descriptions and task requirements like \texttt{``I'm working for the science communication department and writing media posts''}. Additionally, \texttt{J11} asked ChatGPT to provide five ideas as references, thus leveraging the creative abilities of LLMs. Similarly, participant \texttt{J13}, who had the lowest similarity score amounting to 0.783 in the post-training task for \textsc{Article Psy} (subgroup average 0.857), employed a chain-of-thought strategy by prompting with \texttt{``That's a good start. Could you address this post to a kid aged 10?''} Generally, we observe that most journalists adopt different prompting strategies and apply the refinement methods they learned during the training. Those who score lower in similarity employ these techniques more frequently and accurately. 

\section{Discussion}


In this paper, we study the effects of prompt engineering training on user experience (RQ1), text accuracy (RQ2), and reader perception (RQ3) among professional journalists. Our findings reveal that, while our training significantly improves journalists' perceived expertise with LLMs, it has a mixed impact on the perceived helpfulness of LLMs. The effect of training on the accuracy of generated content varied depending on the complexity of the scientific article. Lastly, we find a nuanced impact of training on reader perception across different text quality dimensions.


\textbf{Practical implications:} Our findings suggest that prompt engineering training can effectively enhance users' perceived expertise in interacting with LLMs, though it may not uniformly improve all aspects of their experience, such as perceived helpfulness. For companies considering the integration of LLMs into their workflows, this raises the question of how prompt engineering training can be designed so that they improve both the perceived expertise in interacting with LLMs and the perceived helpfulness of LLMs for the workers. Future research could focus on developing best practices that companies can use to maximize the benefits of prompt engineering training to ensure it meets the needs of users with varying levels of expertise.


The mixed outcomes in user experience, particularly the decreased perceived helpfulness after training, suggest that current LLM applications may not fully align with user expectations, especially for complex tasks. This points to a critical need within the field of human-computer interactions to develop better auto-prompting systems and user interfaces that can adapt to the user’s level of expertise and the specific task at hand. By incorporating more intuitive and supportive interface designs, developers can help bridge the gap between user capabilities and the potential of LLMs, making these tools more accessible and effective for a broader range of users.


Lastly, our findings have significant implications for the field of journalism, particularly as LLMs become more integrated into the profession. While LLMs can assist in streamlining content creation, the dissatisfaction with stylistic elements and cultural adaptability point to challenges that journalists face when using LLMs. Journalists may need to develop new skills to effectively collaborate with LLMs to ensure that the content produced meets the high standards expected in the field. Additionally, journalism as a profession may need to engage more critically with LLMs and advocate for tools that better reflect the diverse cultural contexts in which they operate. This could involve closer collaboration between journalists and LLM developers to refine these tools in ways that better serve the unique needs of the profession.

\textbf{Limitations and future research:} While our study provides valuable insights into the impact of prompt engineering training, several limitations present opportunities for future research. First, the relatively small sample size of 29 participants may limit the generalizability of the findings, but it is nevertheless consistent with prior field experiments that involve similar tasks for professionals \cite{Li.2024, Osone.2021, Huang.2020, Senoner.2024}. Second, the study was conducted with journalists from a specific geographic region, which may influence the cultural adaptability issues highlighted in the results. Future research could expand the sample size and include participants from diverse backgrounds to validate and extend these findings. Future research should also explore the long-term effects of prompt engineering training on user experience and output quality, particularly as LLMs continue to evolve. Third, in our study, we use ChatGPT-3.5, which might impact the output quality of the texts written by the journalists with the LLM. Yet, the performance of ChatGPT-3.5 was state-of-the-art at the time of our experiment, and it was the version that was freely available at the time of our study. This makes it highly used by people and it has a low interaction threshold. Fifth, although some post-hoc evaluations did not yield statistically significant results, this could be attributed to the exploratory nature of the study. It is essential to recognize that some dimensions of the outcomes did not show changes, which indicates the need for further investigation into which aspects of prompt engineering are most impactful. 

Further, our work underscores the critical role of teaching good prompting to equip users with the tools to enhance the output quality of LLMs. Unlike studies that compare different prompting strategies, our research focuses on the human aspect of crafting prompts. This highlights an important area for further exploration: how personalized and context-specific prompting strategies can be developed to consistently improve task outcomes across different scenarios. Understanding the nuances of effective prompting is essential for professionals aiming to leverage LLMs for more accurate and relevant results in their work.

\section{Conclusion}

Our work contributes to the existing body of knowledge on how prompt engineering can shape the interaction between professionals and LLMs. As LLMs continue to transform the way professionals create content, it will be increasingly important to equip users with the skills to navigate these technologies effectively.

\begin{acks}
We express our appreciation to the journalists who participated in this study. Their contributions provided valuable insights that enhanced our research.
\end{acks}

\bibliographystyle{ACM-Reference-Format}
\bibliography{literature}

\appendix

\section{Scientific articles used in the field experiment}
\label{appendix:articles_abstract}

Here are the two article summaries for the writing tasks during the field experiment.

\subsection{\textsc{Article Psy}}

\emph{The following abstract is quoted from \cite{Ross.2024}:}

\noindent\fbox{%
    \parbox{\textwidth}{%
\footnotesize

\begin{center}
    \textbf{Estimated Average Treatment Effect of Psychiatric Hospitalization in Patients With Suicidal Behaviors. \\A Precision Treatment Analysis}
\end{center}

\textbf{Importance:} Psychiatric hospitalization is the standard of care for patients presenting to an emergency department (ED) or urgent care (UC) with high suicide risk. However, the effect of hospitalization in reducing subsequent suicidal behaviors is poorly understood and likely heterogeneous.

\textbf{Objectives:} To estimate the association of psychiatric hospitalization with subsequent suicidal behaviors using observational data and develop a preliminary predictive analytics individualized treatment rule accounting for heterogeneity in this association across patients.

\textbf{Design, Setting, and Participants:} A machine learning analysis of retrospective data was conducted. All veterans presenting with suicidal ideation (SI) or suicide attempt (SA) from January 1, 2010, to December 31, 2015, were included. Data were analyzed from September 1, 2022, to March 10, 2023. Subgroups were defined by primary psychiatric diagnosis (nonaffective psychosis, bipolar disorder, major depressive disorder, and other) and suicidality (SI only, SA in past 2-7 days, and SA in past day). Models were trained in 70.0\% of the training samples and tested in the remaining 30.0\%.

\textbf{Exposures:} Psychiatric hospitalization vs nonhospitalization.

\textbf{Main Outcomes and Measures:} Fatal and nonfatal SAs within 12 months of ED/UC visits were identified in administrative records and the National Death Index. Baseline covariates were drawn from electronic health records and geospatial databases.

\textbf{Results:} Of 196\,610 visits (90.3\% men; median [IQR] age, 53 [41--59] years), 71.5\% resulted in hospitalization. The 12-month SA risk was 11.9\% with hospitalization and 12.0\% with nonhospitalization (difference, $-$0.1\%; 95\% CI, $-$0.4\% to 0.2\%). In patients with SI only or SA in the past 2 to 7 days, most hospitalization was not associated with subsequent SAs. For patients with SA in the past day, hospitalization was associated with risk reductions ranging from $-$6.9\% to $-$9.6\% across diagnoses. Accounting for heterogeneity, hospitalization was associated with reduced risk of subsequent SAs in 28.1\% of the patients and increased risk in 24.0\%. An individualized treatment rule based on these associations may reduce SAs by 16.0\% and hospitalizations by 13.0\% compared with current rates.

\textbf{Conclusions and Relevance:} The findings of this study suggest that psychiatric hospitalization is associated with reduced average SA risk in the immediate aftermath of an SA but not after other recent SAs or SI only. Substantial heterogeneity exists in these associations across patients. An individualized treatment rule accounting for this heterogeneity could both reduce SAs and avert hospitalizations.
}}

\subsection{\textsc{Article GunViolence}}

\emph{The following abstract is quoted from \cite{Semenza.2024}:}

\noindent\fbox{%
    \parbox{\textwidth}{%
\footnotesize

\begin{center}
    \textbf{Gun Violence Exposure and Suicide Among Black Adults}
\end{center}

\textbf{Importance:} Black individuals are disproportionately exposed to gun violence in the US. Suicide rates among Black US individuals have increased in recent years.

\textbf{Objective:} To evaluate whether gun violence exposures (GVEs) are associated with suicidal ideation and behaviors among Black adults.

\textbf{Design, Setting, and Participants:} This cross-sectional study used survey data collected from a nationally representative sample of self-identified Black or African American (hereafter, Black) adults in the US from April 12, 2023, through May 4, 2023.

\textbf{Exposures:} Ever being shot, being threatened with a gun, knowing someone who has been shot, and witnessing or hearing about a shooting.

\textbf{Main Outcomes and Measures:} Outcome variables were derived from the Self-Injurious Thoughts and Behaviors Interview, including suicidal ideation, suicide attempt preparation, and suicide attempt. A subsample of those exhibiting suicidal ideation was used to assess for suicidal behaviors.

\textbf{Results:} The study sample included 3015 Black adults (1646 [55\%] female; mean [SD] age, 46.34 [0.44] years [range, 18--94 years]). Most respondents were exposed to at least 1 type of gun violence (1693 [56\%]), and 300 (12\%) were exposed to at least 3 types of gun violence. Being threatened with a gun (odds ratio [OR], 1.44; 95\% CI, 1.01--2.05) or knowing someone who has been shot (OR, 1.44; 95\% CI, 1.05--1.97) was associated with reporting lifetime suicidal ideation. Being shot was associated with reporting ever planning a suicide (OR, 3.73; 95\% CI, 1.10--12.64). Being threatened (OR, 2.41; 95\% CI, 2.41--5.09) or knowing someone who has been shot (OR, 2.86; 95\% CI, 1.42--5.74) was associated with reporting lifetime suicide attempts. Cumulative GVE was associated with reporting lifetime suicidal ideation (1 type: OR, 1.69 [95\% CI, 1.19--2.39]; 2 types: OR, 1.69 [95\% CI, 1.17--2.44]; $\geq$3 types: OR, 2.27 [95\% CI, 1.48--3.48]), suicide attempt preparation ($\geq$3 types; OR, 2.37; 95\% CI, 2.37--5.63), and attempting suicide (2 types: OR, 4.78 [95\% CI, 1.80--12.71]; $\geq$3 types: OR, 4.01 [95\% CI, 1.41--11.44]).

\textbf{Conclusions and Relevance:} In this cross-sectional study, GVE among Black adults in the US was significantly associated with lifetime suicidal ideation and behavior. Public health efforts to substantially reduce interpersonal gun violence may yield additional benefits by decreasing suicide among Black individuals in the US.

}}

\section{Oral Interviews}
\label{appendix:oral_interview_records}

Here are additional findings from the oral interviews that we conducted after the field experiment. We focus on some notable impressions from the interviews with the journalists.

\textbf{Question: What is your experience with ChatGPT?}

\textbf{Responses:}
\begin{itemize}
    \item \emph{``It is good for sorting information, but the style is very low (secondary school dissertation). The style and originality are not high-level.''}

    \item \emph{``Speed up really much, but the outcome is not up to expectation. We could not convert to the final 100\% outcome that we would like.''}

    \item \emph{``When it does not listen to the commands, it makes the journalist frustrated and reduces creativity. It burns the energy and makes us ask, do I really want to work with it? At this stage, they are not trustable.''}

    \item \emph{``Concerns: These tools are developed in the US, and they are not necessarily adapted to Europe, especially Swiss culture.''}
\end{itemize}

\textbf{Findings:} Overall, the participants expressed dissatisfaction with the style of the LLM outputs and indicated that the LLM did not achieve the results as desired. The LLM was primarily used to generate initial drafts. Additionally, one participant raised a concern that relying on such technology might reduce cognitive exercise, potentially weakening mental acuity.

\end{document}